\documentclass{PoS}
\newcommand*{\ttbar}{$t\overline{t}$~}
\newcommand*{\alfa}{$\alpha_s^{ISR}$~}
\newcommand*{\hdamp}{$h_{damp}$~}
\newcommand{\MG} {\textsc{mg5}\_a\textsc{mc@nlo}~}
\newcommand{\PHEG} {\textsc{powheg v2}}
\newcommand{\PYTHIA}{\textsc{pythia8}~}
\newcommand{\MCATNLO} {a\textsc{mc@nlo}~}
\newcommand{\CP}{\textsc{cuetp8m2t4}~}
\newcommand{\CO}{\textsc{cuetp8m1}~}
\newcommand{\HERWIG}{\textsc{herwig++}~}
\newcommand{\EE}{\textsc{ee5c}~}
\usepackage{url}
\usepackage[lofdepth,lotdepth]{subfig}

\title{Top quark modelling and generators in CMS}

\ShortTitle{Top quark modelling in CMS}

\author{\speaker{Efe Yazgan on behalf of the CMS Collaboration}\\
        Chinese Academy of Sciences, Institute of High Energy Physics, Beijing, China\\
        E-mail: \email{efe.yazgan@cern.ch}}

\abstract{
Recent top quark event modelling studies done using CMS proton-proton data collected at a centre of mass energies of 8 and 13 TeV and state-of-the-art theoretical predictions accurate to next-to-leading order QCD interfaced with \PYTHIA and \HERWIG event generators are summarised. 
The particle-level top quark (pseudo-top), underlying event measurement in \ttbar events and parton shower tuning using \ttbar events are discussed. 
}

\FullConference{EPS-HEP 2017, European Physical Society conference on High Energy Physics\\
		5-12 July 2017\\
		Venice, Italy}

\begin{document}

\section{Introduction}
Top quark measurements provide important tests of QCD. Better understanding of perturbative and non-perturbative effects is required to obtain the highest possible precision in top quark mass, its interpretation and in other properties. Top quark measurements are also important to improve the accuracy of predictions in different phase space regions in searches for beyond the Standard Model (SM) effects. Precise top quark measurements are also be used  in direct searches for new physics as well, e.g. with effective field theories expanding the SM Lagrangian. 
For these purposes, the uncertainties in the measurements and the predictions need to be at a level where deviations of the predictions of the Monte Carlo (MC) codes or deviations due to new physics effects be visible. State-of-the-art  next-to-leading order (NLO) matrix element (ME) event generators interfaced to new parton shower (PS) codes used in LHC Run II may provide better modelling and eventually reduce the major theoretical uncertainties. 
The theory uncertainties can partially be tested and improved with datasets that allow differential measurements with well-defined top-quark objects. In this note, a selection of recent top quark event modelling studies from CMS\cite{CMS} are discussed.

\section{Particle Level Top Quark}
Simulations at NLO take the finite width of the top quark into account which is important to accurately model the off-shell production of top quarks and their interference with the backgrounds. 
In these calculations the concept of top quark as a particle is not well-defined and MC dependent. One can only use the kinematics of the final-state particles unambiguously. 
A particle-level top quark (also called pseudo-top quark) can be constructed from the final-state objects after hadronisation. Using particle-level top quarks would yield smaller uncertainties from non-perturbative effects and from acceptance corrections because of the similar phase definitions at the particle and detector levels minimising MC dependence. 
The details of particle-level top quark definitions and their adoption in the RIVET \cite{RIVETrep} framework in the official CMS reconstruction code are discussed in \cite{Collaboration:2267573} as a fundamental aspect for current and future measurements of differential production cross sections in both \ttbar and single-top quark production. The results reported in \cite{Collaboration:2267573} indicate that the particle-level top quark definition needs to be optimised depending on the production mode, the final state or the variable and the phase space being investigated. 
 
\section{Underlying event and Parton shower tuning in \ttbar events}
The CMS Run I combination of direct top quark mass measurements at 7 and 8 TeV, in lepton+jets, dilepton, and all-hadronic channels yields a precision of 0.3\% \cite{Khachatryan:2015hba}. In this result, the dominant uncertainties are related to the event modelling. Therefore, to improve top quark mass measurements, dedicated measurements and theory studies are required. 
The the b quark from the top quark decay carries the colour flow.  To become colourless, the b quark "connects" with the beam remnants or other coloured final particles produced in the event. A b jet in the final state can be constructed however the uncertainty in the origin of all the final states in the jet results in "odd clusters" (e.g. see \cite{Argyropoulos:2014zoa}). Therefore, it is important to have an accurate description of b quark fragmentation and hadronisation, as well as UE is needed.  

\begin{figure}[ht!]
\begin{center}
\subfloat[]{\includegraphics[width=0.33\textwidth]{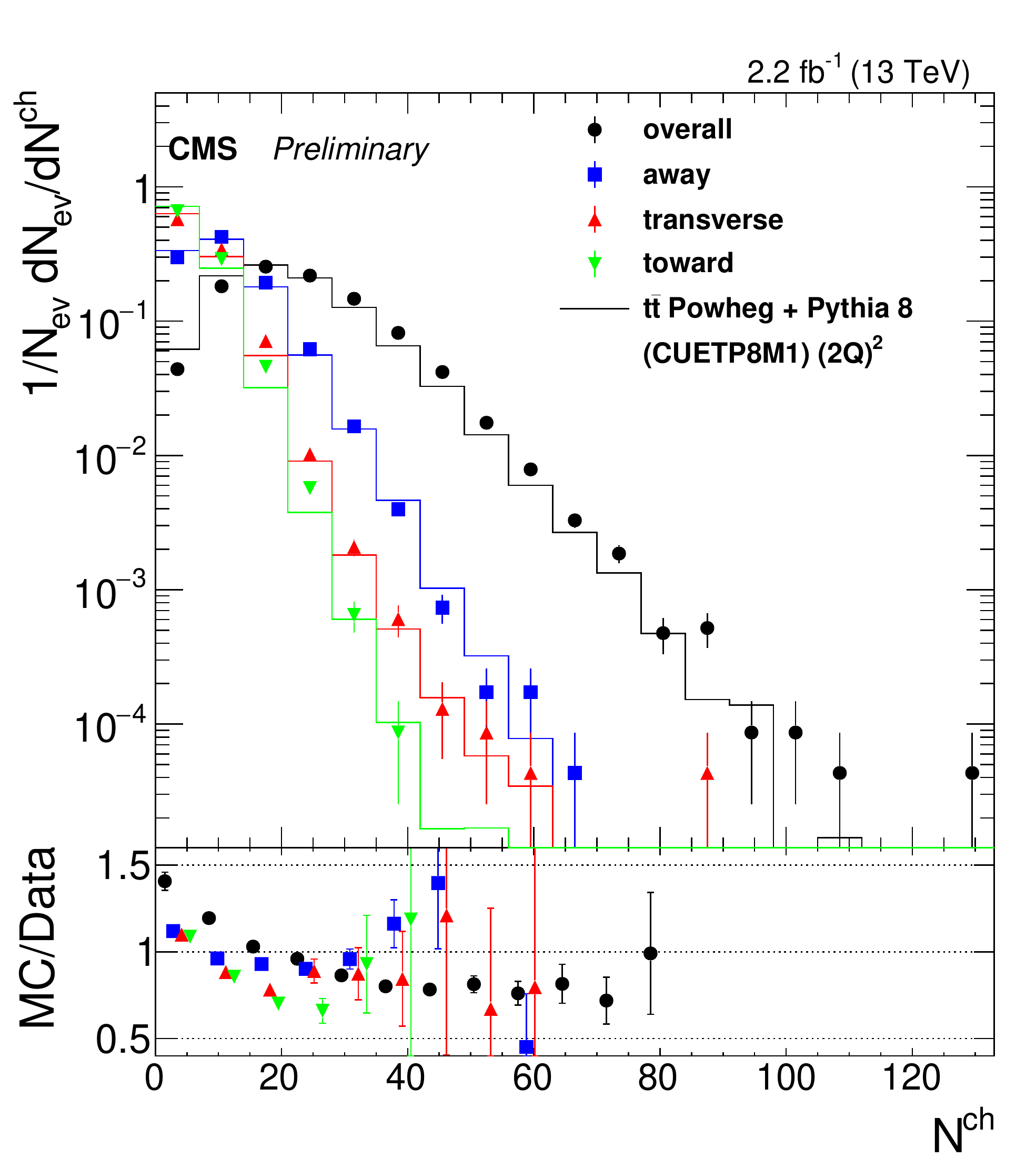}}
\subfloat[]{\includegraphics[width=0.4\textwidth]{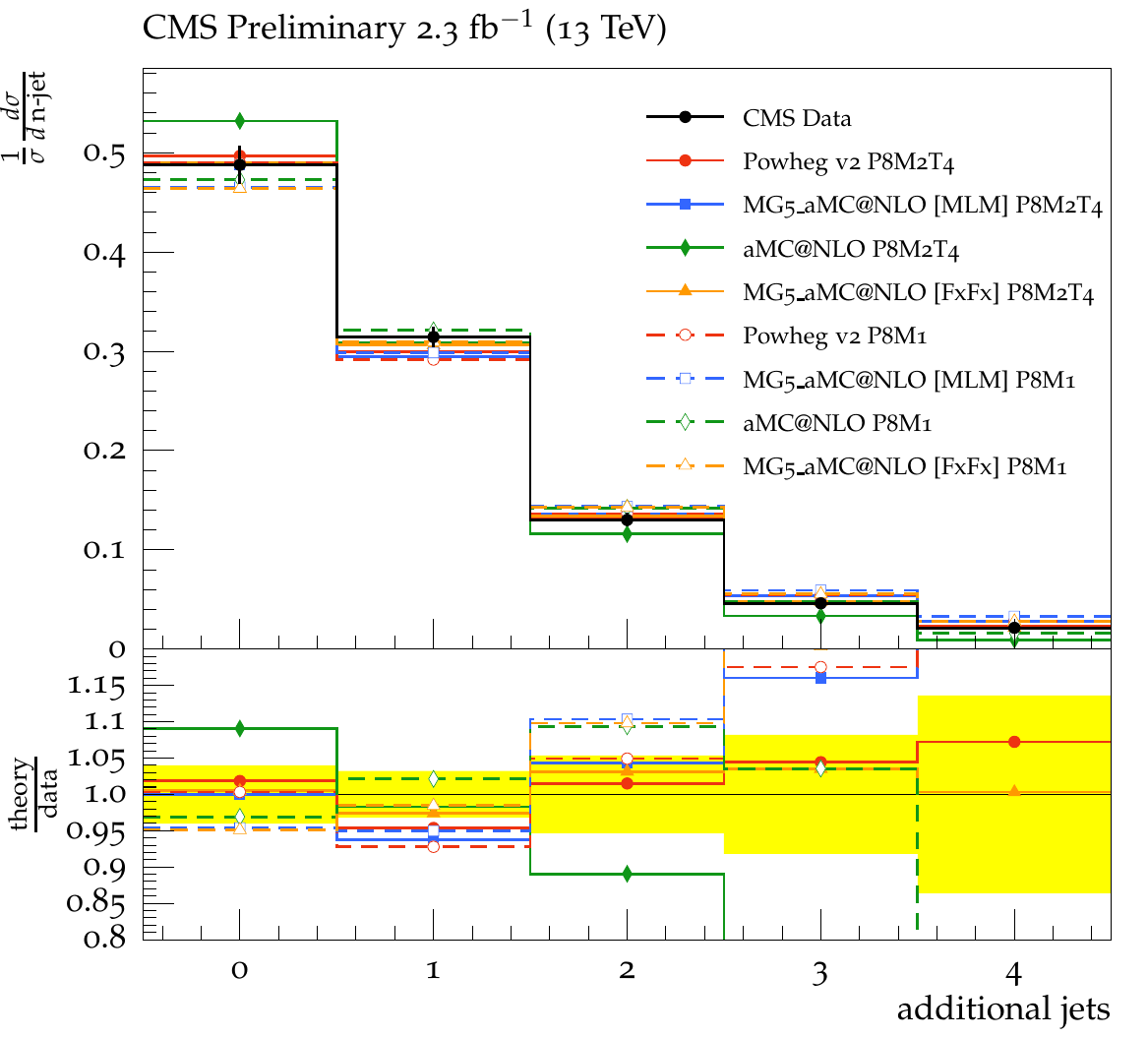}}
\caption{
The charged  particle multiplicity distributions for the away, transverse and toward regions as well as for the overall sample (a). Distributions are obtained with the increased scale, i.e. $(2Q)^2$ that matches the data better. 
The points correspond to the data at the detector level and the lines represent the \PHEG + \PYTHIA predictions with the \CO tune. Each distribution is normalised to one.
CMS data at 13 TeV on the normalised cross section in the lepton+jets channel, as a function of the number of additional jets (b). 
The data are compared with the predictions of \PHEG, \MG, either with MLM matching or FxFx merging, and with \MCATNLO interfaced with \PYTHIA with \CO and \CP tunes. Also shown is the ratio of the theory and the data (theory/data), where the yellow band indicates the total experimental uncertainty of the data.
}
\label{fig:fig}
\end{center}
\end{figure}

UE measurements use the highest $p_T$ charged-particle jet, the highest $E_T$ calorimeter jet, ad the Z-boson direction as the leading-object to define regions of $\eta-\phi$ space, in the toward, away, and transverse regions. The transverse region is particularly sensitive to the modelling of the UE. 
The \PYTHIA tune \CO and the \HERWIG tune \EE \cite{Seymour:2013qka} are reconstructed by fitting the UE data at several centre-of-mass energies, where the leading object is the highest $p_T$ charged particle or the charged-particle jet in the event. These tunes describe well the UE as measured in Z-boson production. However, very little is known about UE in heavy quark production. 
Ref. \cite{CMS:2015usp} compared detector-level top-quark production data at 13 TeV with the \PYTHIA \CO tune and the \HERWIG \EE tune after detector simulation in \ttbar enriched events in the lepton+jets channel. Both of the parton shower models are interfaced with \PHEG. Fair agreement is observed between \PHEG+\PYTHIA \CO tune predictions. It is also observed that UE is sensitive to QCD scales. 
Fig.~\ref{fig:fig}a displays the charged-particle multiplicity when the PS scale is increased from its default value. 
A complete measurement of UE in \ttbar events at particle-level may lead to more precise top quark mass with better understood systematic uncertainties. 

It is observed that the predictions of the NLO MC ME generators + \PYTHIA \CO tune \cite{Khachatryan:2015pea} overshoot the $\sqrt{s}=$ 8 \cite{Khachatryan:2015mva,CMS-PAS-TOP-16-021} and 13 TeV \cite{CMS-PAS-TOP-16-021,Khachatryan:2016mnb} data for large jet multiplicities when out of the box parameters are used, while all other distributions are modelled well (except top quark $p_T$). 
The tune \CO is based on the Monash tune \cite{Skands:2014pea}. 
 Accurate predictions of this observable is particularly important in measurements of the Higgs boson and many new physics search analysis. 
 To improve the description of high jet multiplicities in \ttbar events, a number of parameters have been studied and the most sensitive ones to jet kinematics in \ttbar events are selected and optimised. The strong coupling parameter at $m_Z$ for initial-state radiation in the PS, \alfa, and the \hdamp parameter that controls the jet matching in the \PHEG+\PYTHIA \cite{Nason:2004rx,Frixione:2007vw,Alioli:2010xd} setup are tuned using Run 1 data on jet activity in \ttbar events. The Monash tune for \alfa adopts the $\alpha_s^{FSR}$ value tuned to LEP event shapes. This is found to be the main cause of overproduction of jets in the MC.  We tuned the values of \alfa and \hdamp using the jet multiplicity and leading additional jet $p_T$ distributions in the dilepton final state measured at $\sqrt{s}=8$ TeV using the PROFESSOR tool \cite{Buckley:2009bj}. In this procedure, all other \PYTHIA parameters are kept fixed to the ones in \CO tune.  It is observed that \alfa impacts mostly $N_{jets}>3$, while \hdamp affects the ratio of 2-to-3-jet events and the leading additional jet $p_T$. This is in agreement with the fact that the leading additional jet, in the \PHEG+\PYTHIA configuration, stems from the real radiation calculated by the \PHEG~ generator. The tuning procedure yields \hdamp=$1.581^{+0.658}_{-0.585}\times m_t$ and \alfa=0.1108$^{+0.0145}_{-0.0142}$. The tuned \alfa value agrees with the PDG value of $\alpha_s(M_Z)=0.1181\pm0.0011$ \cite{pdg} well within uncertainties. Fixing \hdamp to its default value of $m_t$, a re-tuning of \alfa alone to the same data yields \alfa=0$.115^{+0.021}_{-0.019}$ \cite{supptop12041} again in agreement with the PDG value. \PHEG+\PYTHIA with these optimised parameters cures the overshoot of the \CO at high jet multiplicities. 
  
 The jet activity mainly constrains those parameters that control the probability for parton emission and the interplay between hard and soft parton emission. The jet activity, however, does not strongly constrain the global production of hadrons known as the underlying event (UE). Therefore, \alfa as determined from \ttbar jet kinematics can be used as a fixed parameter to tune the UE. See ref. \cite{CMS-PAS-TOP-16-021} for the details of the \CP tune derived fixing \alfa to 0.1108.  The performance of the \CP \PYTHIA tune is evaluated in different configurations. It is found that both \PHEG+\PYTHIA and \MG+ \PYTHIA with FxFx merging \cite{Frederix:2012ps}   describe the top quark data well (except for top quark $p_T$ independent of the tune), while \MG+\PYTHIA with MLM matching \cite{Alwall:2007fs} and the inclusive \MCATNLO+\PYTHIA does not describe the data in general (e.g. see Fig.~\ref{fig:fig}b). It is also observed that the global event variables such as $H_T$ or $S_T$ do not get modified significantly with the change of \alfa (except for \MG + \PYTHIA 
 [MLM] and \MCATNLO+\PYTHIA independent of the tune for some variables such as the jet multiplicity) \cite{CMS-PAS-TOP-16-021}. The comparisons of predictions of \PHEG+\PYTHIA with the \CP tune for six different differential  cross-sections to the ones measured with  35.9 $fb^{-1}$ data at $\sqrt{s}=13$ TeV and yield an overall p-value from $<0.01$ when theory uncertainties are ignored to 0.91 when theory uncertainties are included \cite{CMS:2017uda}. 

\end{document}